Dynamic origin-to-destination routing of wirelessly connected, autonomous vehicles on a congested network.

L.C. Davis

10244 Normandy Dr., Plymouth, MI 48170, United States

**Abstract**

Up-to-date information wirelessly communicated among vehicles can be used to select the optimal route between a given origin and destination. To elucidate how to make use of such information, simulations are performed for autonomous vehicles traveling on a square lattice of roads. All the possible routes between the origin and the destination (without backtracking) are of the same length. Congestion is the only determinant of delay. At each intersection, right-of-way is given to the closest vehicle. There are no traffic lights. Trip times of a subject vehicle are recorded for various initial conditions using different routing algorithms. Surprisingly, the simplest algorithm, which is based on the total number of vehicles on a route, is as good as one based on computing travel times from the average velocity of vehicles on each road segment.

Highlights

- Congestion of autonomous vehicles travelling on a square lattice of roads is simulated.
- Wireless communication among all vehicles is assumed.
- Optimal routes from an origin to a destination are found using different algorithms.

1. Introduction

Wirelessly connected, autonomous vehicles offer the potential for better vehicle safety and improved traffic flow, among other benefits. [1-8] From a research perspective, these systems present an interesting challenge to determine how to exploit them in an optimal manner. One such topic is to find the fastest path between a given origin (O) and destination (D) amongst multiple routes with time varying congestion. Many papers [9-12] have been published on the "OD" problem, frequently with the emphasis on mathematical techniques, such as the Dijkstra algorithm [13], to find the fastest route in a computationally efficient way. In the presence of congestion, the fastest route might not be the shortest route. Researchers tend to use measurements of flow to estimate travel time on individual road segments. A common practice is to use a function (travel time as a function of flow) from the US Bureau of Public Roads. [14] The present paper is a report of simulations to determine the fastest OD route for autonomous vehicles with wireless connections navigating through congestion of their own making.

Wireless communication among vehicles can provide useful information. The simplest possibility is just broadcasting the number of vehicles on each link. The average velocity of vehicles on any link could also be provided. In this paper I consider how information from connected vehicles on a road network can be used to predict the optimal route for a given OD. The network studied is a square lattice of identical links (single lane, one way), so each route to be analyzed is the same length (assuming no north-bound to west-bound or east-bound to south-bound turns). Congestion is the only factor that determines travel time. Vehicles are taken to be identical, autonomous cars equipped with adaptive cruise control, which is consistent with the assumption of an ideal wireless connection. [15] The network is simple enough that enumerating paths is straightforward. The emphasis of this research is to use information from wireless connections about realistic congestion caused by many vehicles moving on the network, rather than a surrogate like flow at an intersection, when attempting to find optimal routes.

The model is described in Sec. 2 of the paper. Simulations for different algorithms are compared in Sec. 3. The nature of the congestion through which the subject vehicle travels is also described in Sec. 3. Conclusions and discussion are presented in Sec. 4.

## 2. The model

The goal is to discover routing algorithms that reduce the travel time from a given origin to a given destination in the presence of traffic congestion. Each vehicle is taken to be autonomous with wireless connection to the subject vehicle, whose trip times from origin to destination are recorded. Congestion occurs on the network of roadways when sufficient numbers of vehicles are present. Except for the subject vehicle, all vehicles travel on the roads with a non-zero probability to turn randomly at intersections. The road network is a square lattice of single-lane, one-way roads (Fig. 1); each link is of the same length $N_c D$ where $N_c$ is an integer and $D$ is the minimum distance between vehicles at zero velocity. The east-west roads are labelled by $L$ and the north-south roads are labelled by $K$. Odd values of $L$ denote roads running east and even values west. Likewise odd values of $K$ indicate north and even south. The subject vehicle travels from the lower left to the upper right corner of the lattice using only east bound and north bound roads. Travel time begins when the subject vehicle crosses intersection ($K$, $L$) = (1, 1) and ends when it reaches a point east of ($K_{max}$, $L_{max} - 1$). I take $K_{max} = L_{max}$ to be even integers. The boundary conditions are periodic so that the number of vehicles is conserved.

East-bound vehicles obey the following equations of motion:

$$\frac{dv_i}{dt} = min\{a_{accel}, a_i^d\}, \tag{1a}$$

$$a_{accel} = 1 \text{ m/s}^2, \tag{1b}$$

$$a_i^d = \alpha \left[\frac{x_{lead} - x_i - D}{h_d} - v_i\right] + k_d(v_{lead} - v_i). \tag{1c}$$

Here "*lead*" refers to the vehicle immediately in front of vehicle $i$ or it can be a stopping point at an intersection if vehicle $i$ does not have the right of way, which is determined by the closest vehicle approaching the intersection. Note that at each intersection there is only one direction a vehicle can turn because the roads are one way; also there are no traffic lights in this model. Additionally, other complications such as lane changes or overtaking are not considered in this model. Vehicle velocities are limited to [0, $v_{lim}$]. Analogous equations apply to vehicles moving in other directions. The update time increment is 0.1 s. Parameter values for simulations are $\alpha$ = 2/s, $h_d$ = 1 s, $k_d$ = 1/s, $D$ = 7.5 m, and $v_{lim}$ = 32 m/s.

Eq. (1c) is a common form of the control algorithm for ACC that uses feedback from the headway error $x_{lead} - x_i - D - h_d v_i$ and the relative velocity error $v_{lead} - v_i$. Acceleration is limited to a comfortable value of 1 m/s². The values of parameters are typical of previous analyses of the dynamics of platoons and ensure string stability. See Refs. [16-21].

In the present work, traffic lights were not considered because wirelessly connected vehicles do not necessarily require them. However, if manual vehicles comprise part of the traffic, their effects are important. Traffic light timing and phasing, vehicle speeds and other factors determine route times and green-wave paths. These effects have been analyzed by Nagatani and coworkers. See Refs. [22, 23].

The periods for the periodic boundary conditions are

$$X_{period} = (K_{max} + 2)N_c D \qquad (2)$$

for east-west roads and

$$Y_{period} = (L_{max} + 2)N_c D \qquad (3)$$

for north-south roads. Note that in the region $K_{max}N_c D < x < X_{period}$ there are no cross roads to cause congestion and similarly for $L_{max}N_c D < y < Y_{period}$. The destination of the subject vehicle is $x = (K_{max} + 2)N_c D, y = (L_{max} - 1)N_c D$. The origin is at $x = 0, y = N_c D$.

Initially $N_v(L)$ vehicles are evenly spaced on the east-west road $L$ and $M_v(K)$ on the north-south road $K$. The initial velocity of each vehicle is $v_{lim}$. The initial number of vehicles are given by $N_v(L) = N_0 Rnd$ or $M_v(K) = N_0 Rnd$ where $Rnd$ is a random number in [0,1] (different for each road).

One vehicle, called the subject vehicle, travels from O to D in a presence of other vehicles that can cause congestion. At each intersection a vehicle (other than the subject vehicle) attempts to turn with probability $p_{turn}$ whenever two cells (length D) on each side of intersection and the intersection itself (also of length D) are vacant on the new road. For the convenience of following a prescribed route, the subject vehicle requires only one cell on each side of intersection to be vacant. It has mandatory turns at $x = (K_{max} - 1)N_c D$ when east bound and at $y = (L_{max} - 1)N_c D$ when north bound. At other intersections, the route algorithm dictates whether or not the subject vehicle turns. The subject vehicle is chosen as the one closest to $x = 0$ on the $L = 1$ road at $t = T_{rest}$.

Two global algorithms are considered so that the predicted optimum path corresponds to the minimum value of

$$N_{path} = \Sigma_s N_s \tag{4}$$

or

$$T_{path} = \Sigma_s N_c^{seg} D/v_s^{ave} \tag{5}$$

where the sum is over the segments of the path. Note that a segment is two adjacent east-bound or north-bound links (more precisely, a segment includes the intersection in the middle, but not the two on the ends) and $N_c^{seg} = 2N_c - 1$. The subject vehicle does not turn at the intervening intersection because it would be to the south (west) for east (north) bound segments. $N_s$ is the number of vehicles on segment $s$ and $v_s^{ave}$ is the average velocity of vehicles on the segment. In case of no vehicles on s, $v_s^{ave} = v_{lim}$. If the algorithm is static (denoted as "one time"), the optimum route is determined initially at the time the subject vehicle is at the first intersection (1,1). If the algorithm is dynamic, the route is recalculated for the remaining portion of the trip as the subject vehicle arrives at each odd intersection. I also consider a default route which goes from (1,1) to ($K_{max} - 1$,1) to ($K_{max} - 1, L_{max} - 1$) to the destination. See Fig. 1. Cells are used to keep track approximately of where vehicles are, but are *not* used in the calculation of vehicle motion as in a cellular automaton simulation.

3. Simulations

Simulations were performed for various route algorithms. For each value of $N_0$, 100 realizations (random number seeds) were done and the trip times recorded. Trip time is the time the subject vehicle takes to go from the intersection (1,1) to the destination. The length of each link (distance between intersection centers) is 750 m ($N_c = 100$ and $D$=7.5 m) and $L_{max} = K_{max} = 10$. Thus the length of any route is 13.5 Km and it takes 422 s to complete when no congestion exists. The probability $p_{turn}$ = 0.5. The subject vehicle is identified at $T_{rest}$ = 500 s as the western-most vehicle (smallest $x$) on the $L = 1$ road and must be west of the intersection (1,1).

Trip times $T^V$ for the optimal route determined by the algorithm Eq. (5) (called the V algorithm) are plotted against trip times $T^N$ determined by the algorithm based on Eq. (4) (called the N algorithm) for each initial configuration in Fig, 2. $N_0 = 400$. The dashed line is the equal-value line, around which the data cluster tightly. Each route is updated as the subject vehicle reaches a decision point (odd intersection) — thus they are dynamic algorithms. The trip times $T^D$ for the default route are compared to $T^N$ in Fig. 3. Most default travel times are significantly larger than those for the optimal routes. The average trip time is 1418 s compared to 545 s for the optimal routes. Similar results are obtained for other values of $N_0$ as shown in Fig. 4. Trip times clearly depend on the number of vehicles on the network, as well as their initial configuration (number of vehicles assigned to each road).

However, for the congestion simulated in Figs. 2 and 3, the results for the dynamic algorithm are not significantly better than for a static algorithm where the route is chosen at the first intersection and not subsequently updated. See Fig. 5. The utility of a dynamic algorithm can be demonstrated if a perturbation (such as a reduced speed limit in a region of the network) is introduced after the subject vehicle passes the first intersection.

Fig. 6 shows the average velocity of all vehicles vs. time for a typical example with $N_0 = 400$. Within 100 s the average velocity has dropped to ~11 m/s and then remains nearly constant. The subject vehicle is chosen after the system has reached a near steady state at 500 s. The approximately constant value of the average velocity depends somewhat on the initial configuration, varying by a few m/s from run to run. In Fig. 7, the average (100 runs) of the average velocity at $t = 500$ s is plotted against $N_0$. Error bars indicate the variability ($\pm$ one standard deviation) due to initial conditions. The average velocity decreases as the number of vehicles increases, indicating increased congestion.

Unexpectedly, the average of the velocity squared (not shown) is also fairly constant after the initial transient even though individual velocities vary from near zero to the speed limit $v_{lim}$. Because each vehicle is identical, this is equivalent to the kinetic energy of the system of vehicles remaining nearly constant. The system is not conservative, however, because kinetic energy is dissipated by individual vehicles when braking. The near constancy of the total kinetic energy further demonstrates that the system has reached a near steady state by 500 s.

That the V algorithm and the N algorithm give similar results can be understood by noting the correlation between the average velocity on a segment $v_s^{ave}$ and the number of vehicles on segment $N_s$ as shown in Fig. 8.

Fig. 9 shows the number of vehicles $N_s$ on east-bound segments and north-bound segments at various times for a typical simulation. The absence of large changes with time indicates why the static algorithm works almost as well as the dynamic algorithm to find the optimal route as depicted in Fig. 5.

Attempts to find a similar correlation between flow on a segment and the number of vehicles, and thus the travel time, were unsuccessful. Likewise, as Fig. 10 shows, there is no useful relationship between segment travel time $T_s = N_c^{seg} D / v_s^{ave}$ and flow, which is given by $\frac{N_s}{T_s}$. The number of cells in a segment is $N_c^{seg} = 2N_c - 1$. The data were calculated from the data used for Fig. 8.

## 4. Conclusions and discussion

Information provided by wirelessly connected vehicles about congestion on any potential route is shown to be useful for determining the route with the minimum travel time from the given origin to destination. In the scenario evaluated in this paper, all routes are equal in length and speed limit. The determining factor for trip times is the extent to which congestion has set in. In congested traffic, simulations demonstrate that choosing the route with the minimum number of vehicles is as good an indicator of the optimum route as using the average velocity of vehicles on each segment of a route to predict travel time. In some respects the N algorithm uses less information but performs as well as the more detailed V algorithm.

If routes are not of the same length, obviously the V algorithm would be the algorithm of choice. However, if some of the potential routes are equal in length, only the route with the fewest vehicles needs to be considered for comparison to other routes (of different lengths). Because counting the number of vehicles on a segment is simpler than computing their average velocity, using a combination of algorithms reduces the complexity of the optimal route determination.

On the network analyzed in this work, updating gives only marginally better results compared to a single choice made when the subject vehicle reaches the first intersection. Algorithms based on the number of vehicles and on the average velocity are equally effective because there is a strong correlation between the two quantities. However, there is essentially no correlation between the flow on a segment and the number of vehicles or travel time. Because the vehicles on a segment are frequently accelerating and decelerating, one cannot expect a fundamental-diagram relationship between flow and density $N_s/(N_c^{seg} D)$ to hold. Hence the use of the travel time-flow function recommended by the US Bureau of Public Roads [14] is not useful for connected autonomous vehicles on the network considered in this paper.

**Figure captions**

1. The square-lattice network of links used in simulations. Each road is a single, one-way lane with no traffic lights at intersections. East bound roads are labelled as: $L = 1,3 \ldots L_{max} - 1$; West bound: $L = 2,4 \ldots L_{max}$; North bound: $K = 1,3 \ldots K_{max} - 1$; South bound: $K = 2,4 \ldots K_{max}$. The subject vehicle travels from the origin in the lower left corner to the destination which is east of intersection $(K_{max}, L_{max} - 1)$. Trip travel time begins when the subject vehicle crosses intersection (1, 1). The distance from the midpoint of one intersection to the next is $N_c D$ where $N_c$ is an integer (100 in simulations) and $D$ = 7.5 m. The default route is indicated by red lines.

2. The subject vehicle's trip times from origin to destination for each initial configuration of vehicles with $N_0 = 400$. The solid circles are the trip times $T^V$ from the V algorithm [Eq. (5)] plotted against $T^N$ from the N algorithm [Eq. (4)]. The route is updated at each decision point, *i. e.*, when the subject vehicle arrives at intersection denoted by (*K, L*) where *K* and *L* are odd. The dashed line is the equal-value line.

3. For each initial configuration of vehicles ($N_0 = 400$), the trip times $T^D$ for the default route are plotted against the trip times $T^N$ for the optimal route given by the N algorithm [Eq. (4)]. The dashed line is the equal-value line. The default route is from the origin to intersection $(K_{max} - 1, 1)$ to intersection $(K_{max} - 1, L_{max} - 1)$ to the destination. See red route in Fig. 1.

4. For each initial configuration of vehicles the trip time $T^D$ for the default route is plotted against the trip time $T^N$ for the optimal route given by the N algorithm [Eq. (4)]. The dashed line is the equal-value line. (a) $N_0 = 500$, (b) $N_0 = 300$, (c) $N_0 = 200$.

5. For each initial configuration of vehicles ($N_0 = 400$), the trip time $T^{1N}$ calculated from the N algorithm without updating is plotted against $T^N$ from the N algorithm with updating.

6. The average velocity of all vehicles as a function of time in a typical simulation for $N_0 = 400$. The computation of the subject vehicle's optimal route from origin to destination is performed after 500 s.

7. The average velocity of all vehicles averaged over a hundred runs (random number seeds) for each value of $N_0$. Error bars represent $\pm$ one standard deviation. The average is computed at $t = 500$ s.

8. The reciprocal of the average velocity of vehicles in each segment vs. the number of vehicles in the segment. Data for each east-bound and north-bound segment is recorded every second during $t = 100$ to $1000$ s for a typical run (one random number seed). $N_0 = 400$. The dashed line is the linear regression $\frac{1}{v_s^{ave}} = 0.0036 N_s + 0.0185$.

9. The number of vehicles on each segment $N_s$ at $t$ =500, 600, 700 and 800 s for a typical run with $N_0 = 400$. Horizontal rows correspond to east-bound and vertical rows to north-bound segments.

10. Segment travel time $T_s = N_c^{seg} D / v_s^{ave}$ vs segment flow, given by $flow_s = \frac{N_s}{T_s}$. The data were calculated from the data used for Fig. 8.

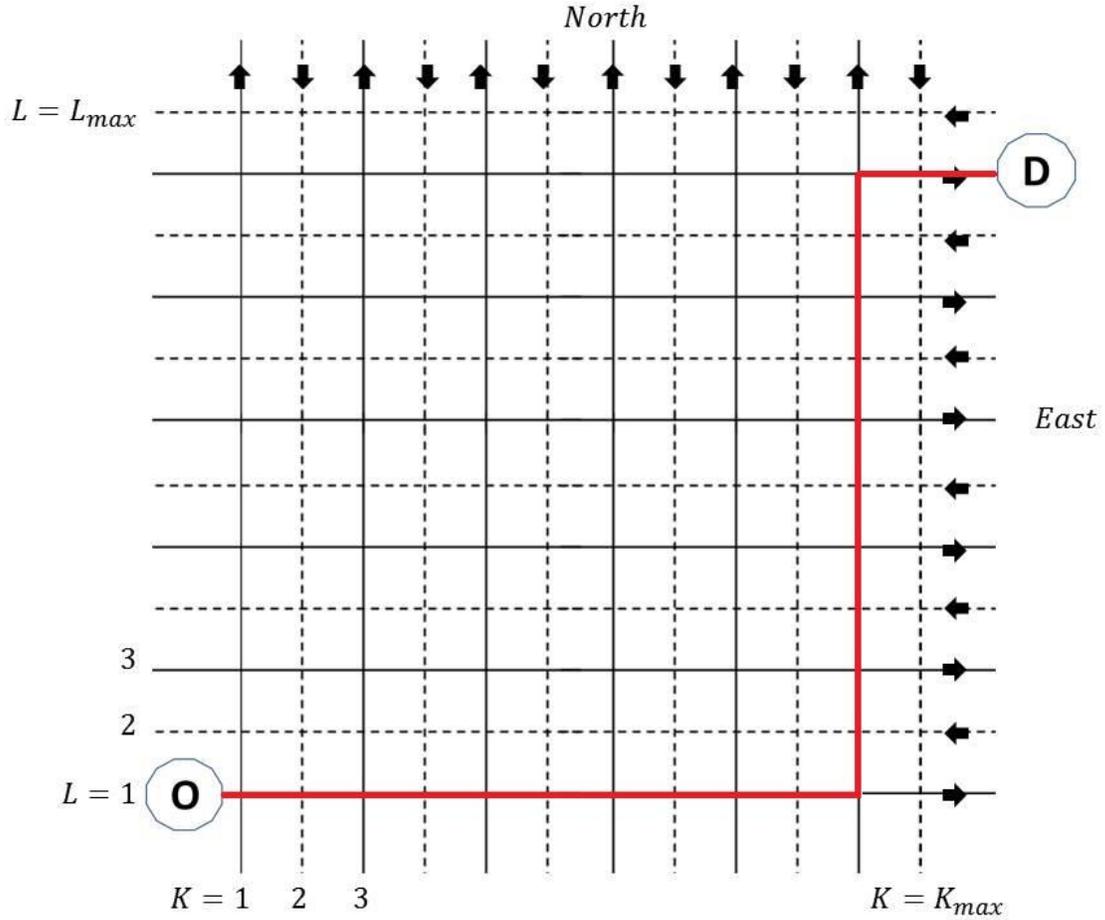

Fig.1. The square-lattice network of links used in simulations. Each road is a single, one-way lane with no traffic lights at intersections. East bound roads are labelled as: $L = 1,3 \ldots L_{max} - 1$; West bound: $L = 2,4 \ldots L_{max}$; North bound: $K = 1,3 \ldots K_{max} - 1$; South bound: $K = 2,4 \ldots K_{max}$. The subject vehicle travels from the origin in the lower left corner to the destination which is east of intersection $(K_{max}, L_{max} - 1)$. Trip travel time begins when the subject vehicle crosses intersection (1, 1). The distance from the midpoint of one intersection to the next is $N_c D$ where $N_c$ is an integer (100 in simulations) and $D$ = 7.5 m. The default route is indicated by red lines.

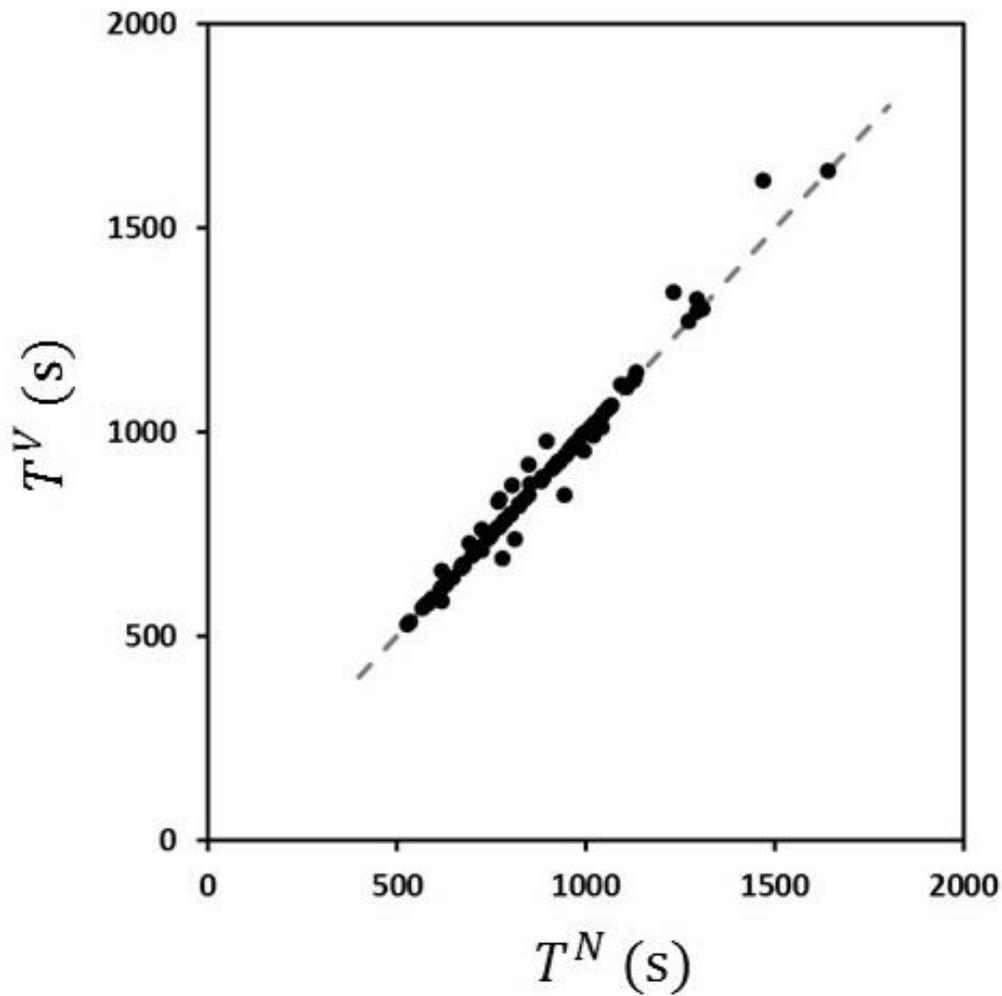

Fig. 2. The subject vehicle's trip times from origin to destination for each initial configuration of vehicles with $N_0 = 400$. The solid circles are the trip times $T^V$ from the V algorithm [Eq. (5)] plotted against $T^N$ from the N algorithm [Eq. (4)]. The route is updated at each decision point, *i. e.*, when the subject vehicle arrives at intersection denoted by (*K*, *L*) where *K* and *L* are odd. The dashed line is the equal-value line.

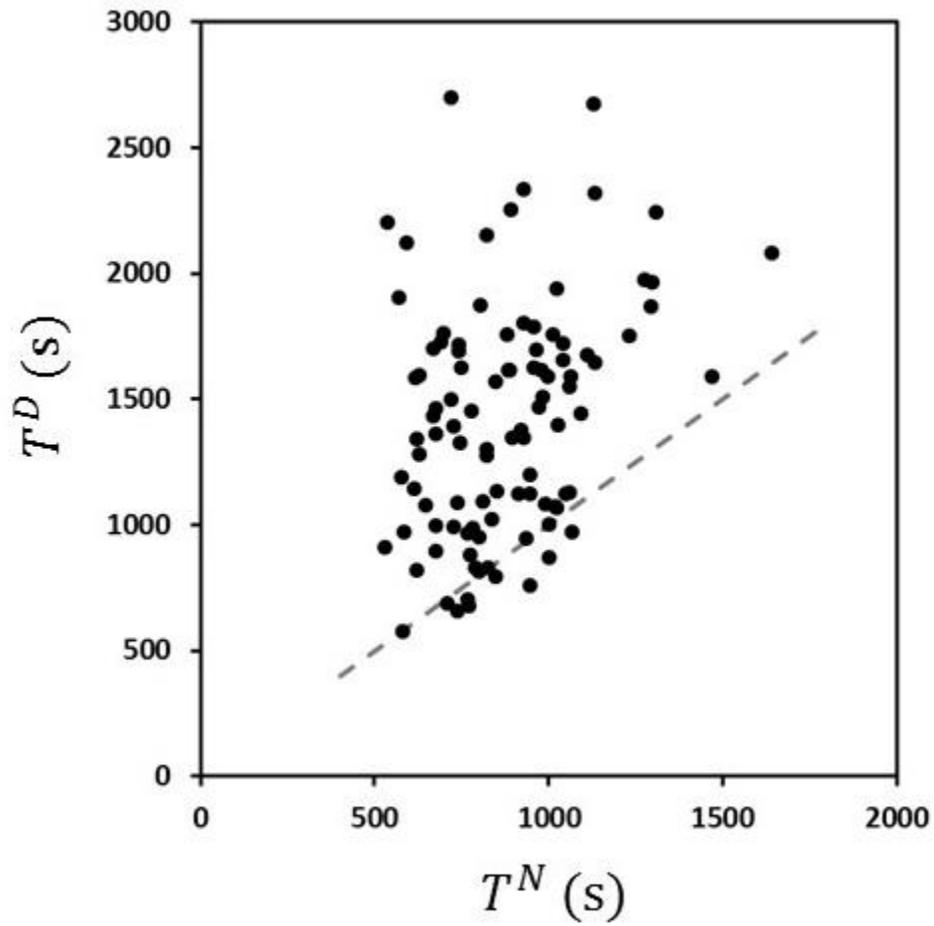

Fig. 3. For each initial configuration of vehicles ($N_0 = 400$), the trip times $T^D$ for the default route are plotted against the trip times $T^N$ for the optimal route given by the N algorithm [Eq. (4)]. The dashed line is the equal-value line. The default route is from the origin to intersection ($K_{max} - 1, 1$) to intersection ($K_{max} - 1, L_{max} - 1$) to the destination. See red route in Fig. 1.

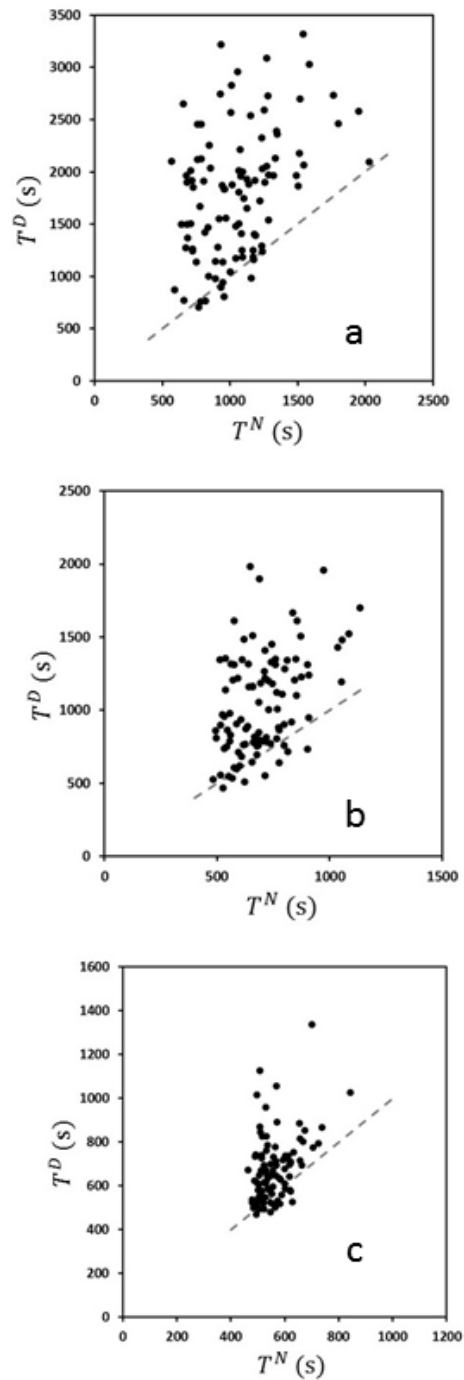

Fig.4. For each initial configuration of vehicles the trip time $T^D$ for the default route is plotted against the trip time $T^N$ for the optimal route given by the N algorithm [Eq. (4)]. The dashed line is the equal-value line. (a) $N_0 = 500$, (b) $N_0 = 300$, (c) $N_0 = 200$.

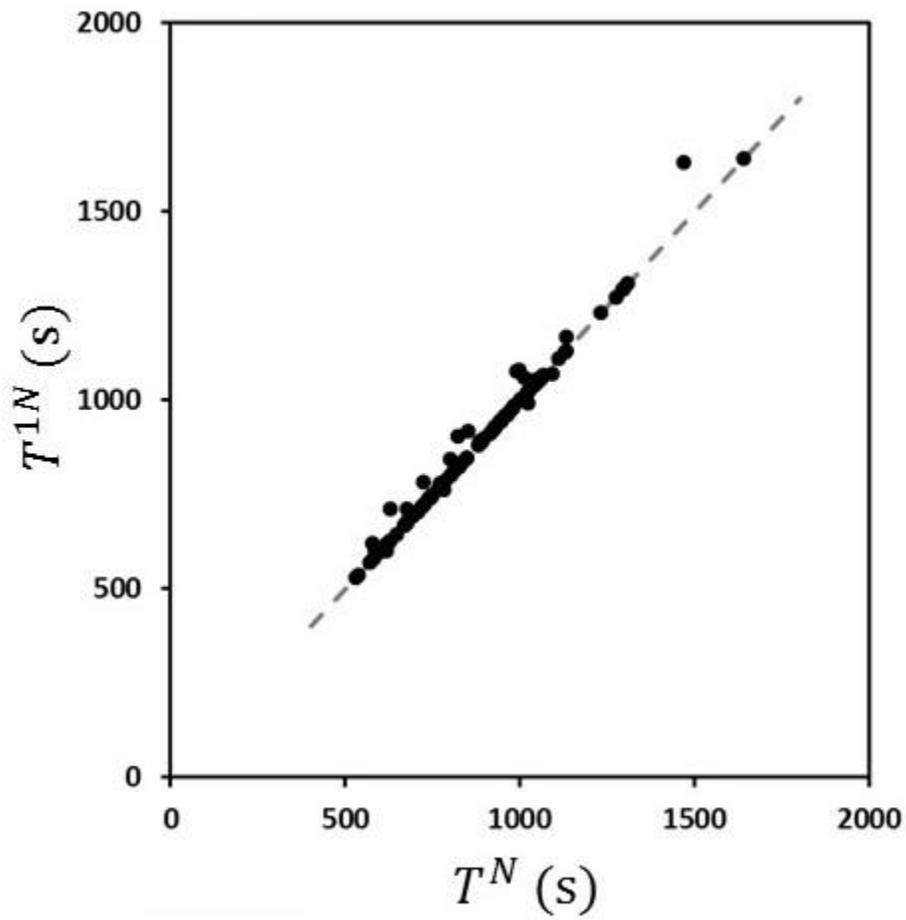

Fig.5. For each initial configuration of vehicles ($N_0 = 400$), the trip time $T^{1N}$ calculated from the N algorithm without updating is plotted against $T^N$ from the N algorithm with updating.

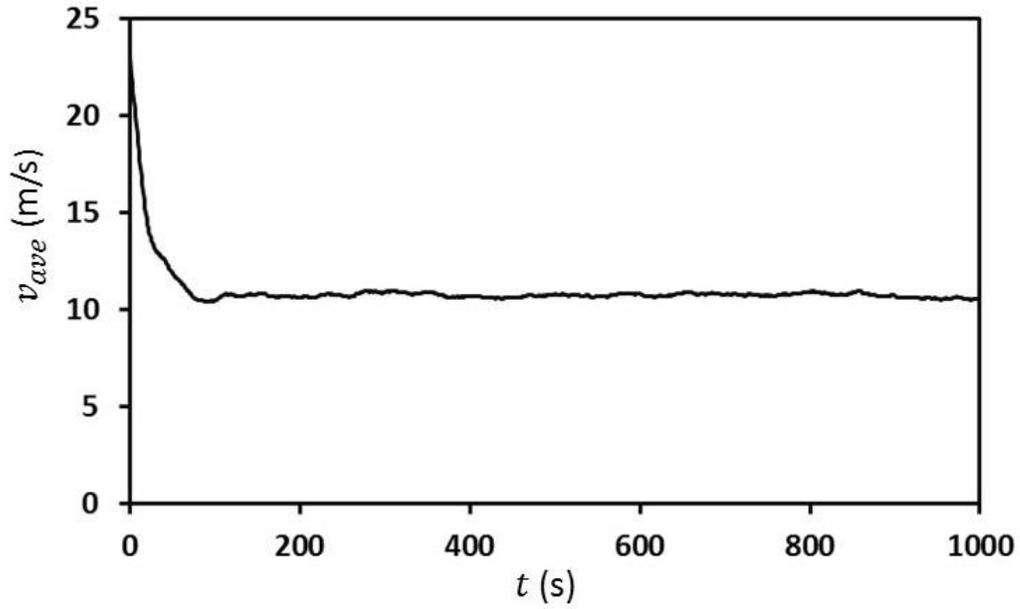

Fig. 6. The average velocity of all vehicles as a function of time in a typical simulation for $N_0 = 400$. The computation of the subject vehicle's optimal route from origin to destination is performed after 500 s.

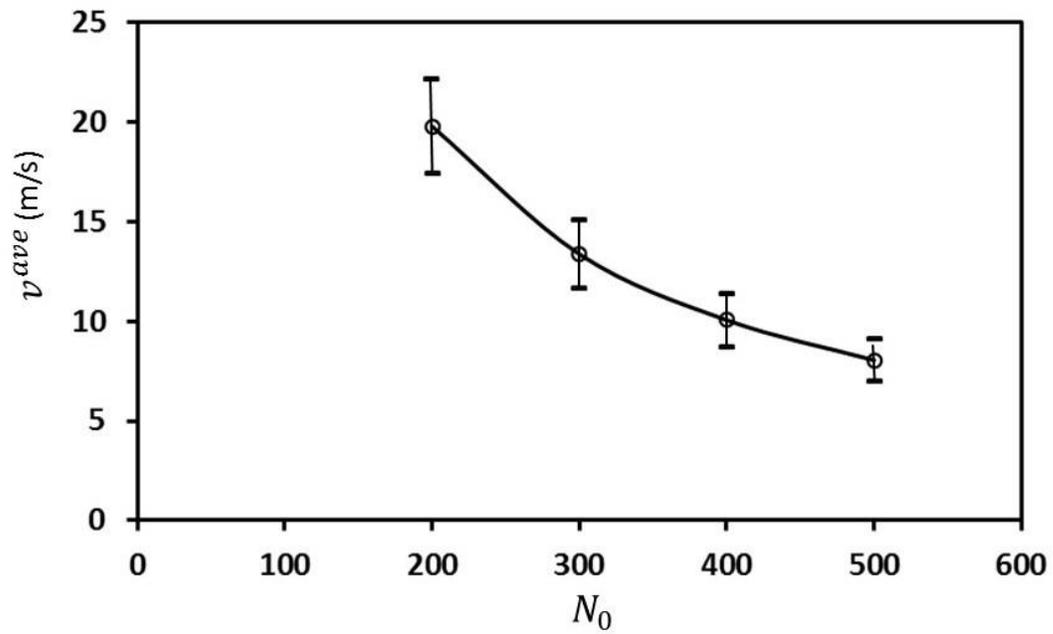

Fig. 7. The average velocity of all vehicles averaged over a hundred runs (random number seeds) for each value of $N_0$. Error bars represent $\pm$ one standard deviation. The average is computed at $t = 500$ s.

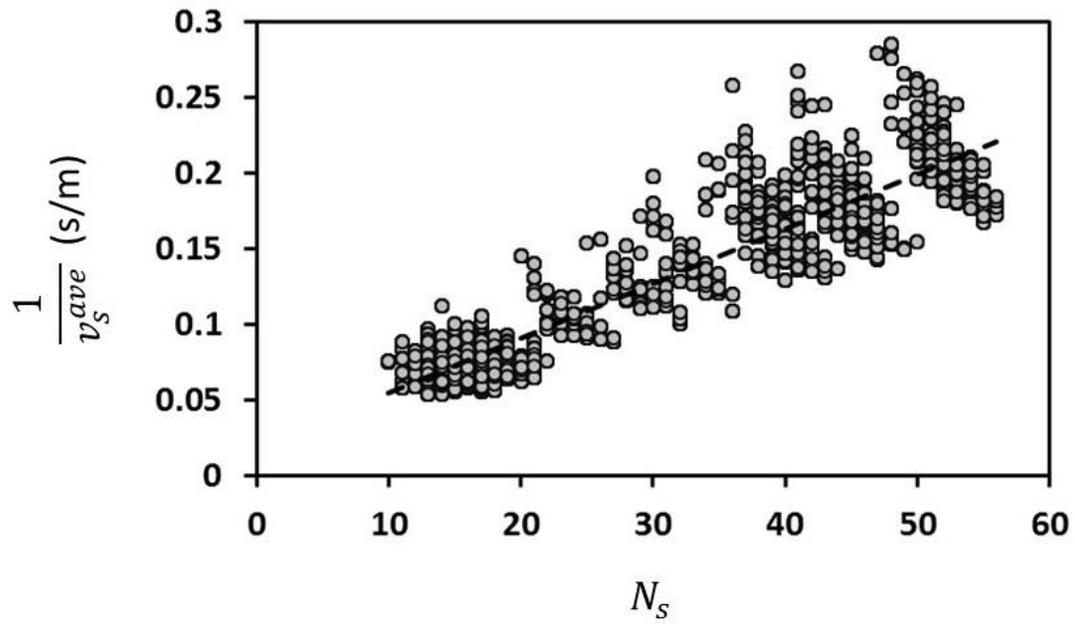

Fig. 8. The reciprocal of the average velocity of vehicles in each segment vs. the number of vehicles in the segment. Data for each east-bound and north-bound segment is recorded every second during $t = 100$ to $1000$ s for a typical run (one random number seed). $N_0 = 400$. The dashed line is the linear regression $\frac{1}{v_s^{ave}} = 0.0036 N_s + 0.0185$.

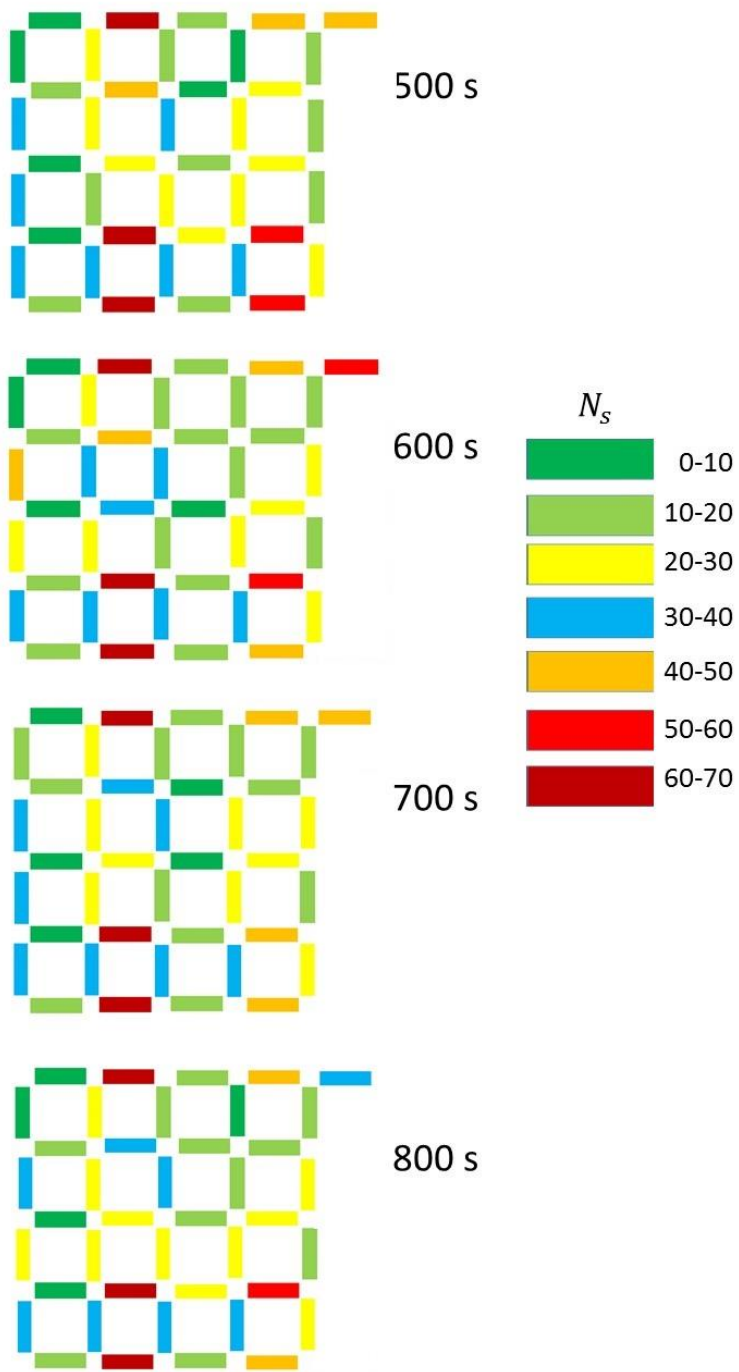

Fig. 9. The number of vehicles on each segment $N_s$ at $t$ =500, 600, 700 and 800 s for a typical run with $N_0 = 400$. Horizontal rows correspond to east-bound and vertical rows to north-bound segments.

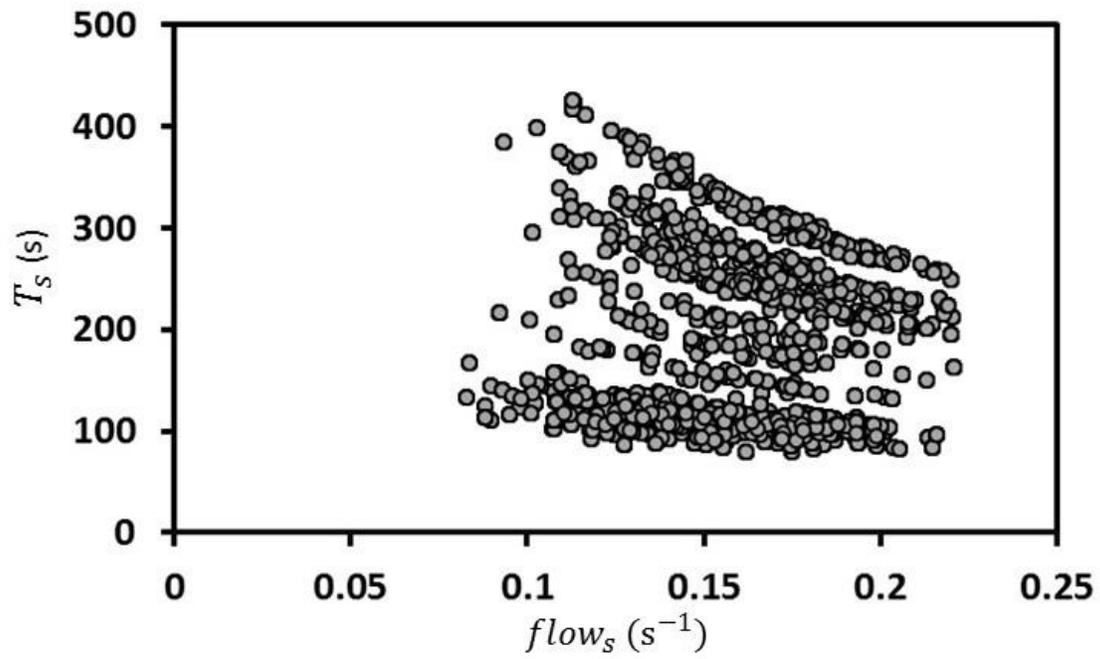

Fig. 10. Segment travel time $T_s = N_c^{seg} D / v_s^{ave}$ vs segment flow, given by $flow_s = \frac{N_s}{T_s}$. The data were calculated from the data used for Fig. 8.